\documentclass[conference]{IEEEtran}
\IEEEoverridecommandlockouts

\usepackage{amsmath, amssymb, amsfonts}
\usepackage{algorithmic}
\usepackage{graphicx}
\usepackage{textcomp}
\usepackage{xcolor}
\usepackage{multirow}
\usepackage{multicol}
\usepackage[utf8]{inputenc}
\usepackage{textgreek}
\usepackage{url}
\usepackage{booktabs}

\usepackage{fancyhdr} 

\fancypagestyle{plain}{
    \fancyhf{} 
    \fancyfoot[C]{\thepage} 
}
\begin{document}
\title{Validating an Instrument for Teachers’ Acceptance of Artificial Intelligence in Education}

\author{
\IEEEauthorblockN{Shuchen Guo}
\IEEEauthorblockA{\textit{School of Education} \\
\textit{Nanjing Normal University}\\
Nanjing, Jiangsu, China \\
shuchen.guo@uga.edu}
\and
\IEEEauthorblockN{Lehong Shi}
\IEEEauthorblockA{\textit{Department of Workforce Education}\\ 
\textit{and Instructional Technology} \\
\textit{University of Georgia}\\
Athens, GA 30602, USA \\
lehong.shi@uga.edu}
\and
\IEEEauthorblockN{Xiaoming Zhai}
\IEEEauthorblockA{
\textit{AI4STEM Education Center} \\
\textit{Department of Mathematics, Science,} \\
\textit{and Social Studies Education} \\
\textit{University of Georgia}\\
Athens, GA 30602, USA \\
xiaoming.zhai@uga.edu}
}

\maketitle

\begin{abstract}

As artificial intelligence (AI) receives wider attention in education, examining teachers’ acceptance of AI (TAAI) becomes essential. However, existing instruments measuring TAAI reported limited reliability and validity evidence and faced some design challenges, such as missing informed definitions of AI to participants. This study aimed to develop and validate a TAAI instrument, with providing sufficient evidence for high psychometric quality. Based on the literature, we first identified five dimensions of TAAI, including perceived usefulness, perceived ease of use, behavioral intention, self-efficacy, and anxiety, and then developed items to assess each dimension. We examined the face and content validity using expert review and think-aloud with pre-service teachers. Using the revised instrument, we collected responses from 274 pre-service teachers and examined the item discriminations to identify outlier items. We employed the confirmatory factor analysis and Cronbach’s alpha to examine the construct validity, convergent validity, discriminant validity, and reliability. Results confirmed the dimensionality of the scale, resulting in 27 items distributed in five dimensions. The study exhibits robust validity and reliability evidence for TAAI, thus affirming its usefulness as a valid measurement instrument.
\end{abstract}


\section{Introduction}
Artificial intelligence (AI) is showing increasing potential to reshape the landscape of future education, forming an emergent research frontier \cite{Guo2024}. This new surge of research focuses on the substantial transformation of AI in teaching, learning, and assessment practices when integrating a wide range of AI applications, such as chatbots, automated scoring systems, and intelligent tutoring systems into education \cite{Chen2022, Guo2024, Zhai2021}. However, realizing this potential is not without challenges.  Successful implementation of AI-based teaching is closely related to factors such as teachers’ perception of and attitudes toward AI, AI literacy, and ethical concerns, etc. Among all the factors, teachers’ willingness to use AI in their teaching has been crucial \cite{Choi2023}. If teachers are reluctant to use AI, this novel technology may be left without truly benefiting any students. This concern is not purely alarmism-- “high access and low use” has constantly been seen in the history of educational technology due to teachers’ low acceptance \cite{AlKanaan2022, AlSubhy2020, Nazaretsky2021}. Can AI be an exception? 

We doubt it. In fact, AI drew substantial concerns when the concept was first introduced several decades ago. People were worried that AI might substitute humans and cause humans to lose control \cite{Li2020,Wang2019}. As AI is growing its autonomy in learning analytics and decision-making and increasingly plays partial roles and responsibilities of teachers (e.g., tutors, teaching assistants), its applications in education draw significant concerns from stakeholders, including teachers, if not more \cite{Jang2022,Li2020}. \cite{Serholt2017}  reported teachers’ fear of AI due to the blurring boundaries between teachers’ roles and AI, which leads to criticism of AI for disrupting teachers’ expertise-based roles. Additionally, the complexity of AI applications and the additional efforts needed to operate them could further reduce teachers’ willingness to use AI \cite{Zhai2021}. In addition to these known facts, pseudo AI harmfulness rumored in the media further upsets teachers and worsens their acceptance of AI \cite{Zhai2024}. Given these concerns, understanding teachers’ acceptance of AI (TAAI) in education has become one of the pivotal factors in this educational transformation process \cite{AlDarayseh2023,Choi2023,SanchezPrieto2017}. 

Some existing studies have assessed TAAI from different aspects, such as including perceived ease of use, perceived usefulness, attitude to AI, and self-efficacy \cite{AlDarayseh2023,Choi2023,Zhang2023}. However, the items used to assess TAAI were primarily limited to revising components of the technology acceptance model (TAM) \cite{Choi2023,Zhang2023}, without comprehensive validation in contexts, resulting in concerns about the psychometric quality and design of items. Meanwhile, the lack of knowledge and awareness of AI among the respondents can also result in inaccurate results, since most of the instruments have not provided participants with an introduction to AI and its use in education. Due to these constraints, previous studies show limitations in understanding teachers’ perspectives on accepting AI in their teaching and identifying the influential factors. Therefore, it is vital to develop a validated instrument for measuring teachers’ acceptance of AI \cite{Zhang2023}. 

To address this gap, we developed and validated an instrument to measure TAAI. We first conceptualized the construct and developed a theoretical framework, followed by developing items and conducting cognitive think-aloud interviews to revise the items. To validate the TAAI, we obtained data from 274 pre-service teachers from diverse discipline backgrounds to test the instrument.  Using item and test analyses, we confirmed TAAI’s validity and reliability. The study answered three research questions: 1) What are the psychometric features of the TAAI items? 2) To what extent are the instrument items reliable in assessing teachers’ acceptance of AI? 3) How valid are the instrument items in assessing teachers’ acceptance of AI?

\section{Challenges of Measuring TAAI}

\subsection{Psychometric Quality}
Although measuring TAAI is an important topic and many studies have reported preliminary findings, the field is suffering from a lack of instruments with robust evidence of psychometric properties. Insufficient psychometric evidence will undermine the methodological rigor of the instrument, in turn significantly compromising the validity of the findings. To obtain a high-quality instrument, researchers must provide psychometric evidence in various aspects to ensure validity. 

A valid instrument needs to be supported by validity evidence, including content, construct, convergent, and discriminant validity. To address the content validity, the developed instrument needs to be put through expert review, pilot test, and cognitive interview; otherwise, the instrument is in lacks evidence of face and content validity \cite{Choi2023,Wang2021}. We have rarely seen research documenting sufficient content validity evidence. With regard to construct validity, studies generally use exploratory factor analysis or confirmatory factor analysis, reporting the fit indices and factor loading to show the factor structure \cite{Zhang2023}. To maintain a high-level construct validity, the number of items in one dimension should be no less than three; otherwise, it could undermine the rigor of the measurement \cite{Chocarro2023}. Furthermore, the presence of items within the same dimension with similar meanings has to be cautious. Overly similar items may be redundant and should be reduced. For example, instruments measuring technology acceptance have integrated “anxiety” as one of the dimensions within TAM and relevant models \cite{SanchezPrieto2017,Wang2021,Zhang2023}. However, the dimension used for measuring anxiety had a small number of items focused on similarly negative feelings that people have with AI, resulting in some redundancy of the instrument. Such item design may yield better reliability results, but it reduces the information that could be obtained from the instrument. Additionally, it can confuse subjects when answering items with similar meanings. 

Moreover, convergent and discriminant validity, as well as item discrimination, matter the instrument’s psychometrical quality. Convergent validity refers to whether items within the same dimension are constructed to represent most of their constructs and eventually converge in the same dimension, while discriminant validity means that each item should be strongly related to its own dimension and weakly related to other dimensions \cite{Wang2024}. The two aspects of validity demand a clear conceptual framework for the instrument. Item discrimination indicates the discriminative power of items to distinguish between respondents with higher and lower scores. To maintain item descriptive, the developed items should be accurate and understandable to the respondents without confusing wording that can impact the appropriate response of subjects \cite{Chatterjee2020}.

However, few studies have taken all the aspects of validity into consideration, leaving the existing instruments of TAAI without robust validity evidence. Other studies that developed instruments to assess teachers’ AI affection with sufficient validity evidence, but fall out of TAAI. For example, \cite{Sanusi2024} developed an instrument to measure pre-service teachers’ AI perception based on the planned behavior theory. They provided evidence about construct reliability, convergent validity, and discriminant validity. However, the instrument is about measuring teachers’ perception of learning AI rather than TAAI, which this study focuses on. 
To sum up, there are limited well-validated instruments to measure TAAI in education to date. The field needs instruments with psychometric quality evidence specifically to measure teachers’ AI acceptance in education.

\subsection{Concerns related to participants’ knowledge of AI in daily life and education}
Participants’ responses would be impacted by their knowledge and understanding of the objects being measured. That is, teachers’ responses to AI acceptance can be impacted by their knowledge and understanding of AI. Without realizing this idea, TAAI instruments may yield invalid conclusions. 

As an emerging technology, AI has been widely used in a variety of applications in our daily lives. However, many people are unaware of the conceptions of AI and unconscious of AI applications, let alone its potential application in education. A survey, conducted in 2022 encompassing 11,004 U.S. adults aimed at gauging people’s awareness of AI applications in daily life found that although 90\% of the participants have heard about the term AI, merely one-third reported having extensive knowledge about AI and can correctly identify all the uses of AI provided in the survey (Pew Research Center, 2023). Similarly, the British Office for National Statistics released an article about public awareness, opinions, and expectations about AI and reported that only 17\% of adults could identify AI usage in daily life \cite{ONS2023}. The findings resonate with a recent study reporting a range of misunderstandings of AI \cite{Bewersdorff2023}. Even though AI has been increasingly introduced into educational contexts, \cite{Alimi2021} found that many pre-service teachers were unaware of AI in their learning. Similarly, many teachers face the same problem of being unfamiliar with AI concepts and its various applications in daily life and education. Without knowing AI concepts and its various applications, teachers might provide incorrect or biased responses when reporting their acceptance and attitude toward AI in education.  

The problem can be partly alleviated by offering stimuli, such as reading materials or video clips at the very beginning of the survey. The approach is considered to be useful in helping participants recall their memories of using technology and get a better understanding of the technology,  thus eliciting more accurate responses during the test \cite{Choi2023,Kim2018,Xu2020}. Studies on the acceptance of technology that involve people with limited knowledge and familiarity with the specific technology have used this methodology \cite{Choi2023}. For example, to evaluate children’s perceptions of conversational agents (CA), \cite{Xu2020} allowed children to interact with the CA during the study, providing them with proximal and in-person experiences to better elicit their perceptions. Kim et al. \cite{Kim2018} provided reading articles to students when investigating their perceptions of AI teaching assistants, given these assistants are relatively new and have not yet been widely implemented. \cite{Choi2023} provided reading material on their educational artificial intelligence tools (EAIT) to support teachers’ understanding of the concept of EAIT before evaluating their acceptance of EAIT. Additionally, \cite{Chocarro2023} provided images of conversations with chatbots before surveying teachers’ attitudes towards chatbots in education. To be noted, as an emerging technology, AI has not been widely spread and integrated in classrooms, thus limiting teachers’ knowledge of AI in education. Therefore, providing stimuli is critical for obtaining reliable results when measuring TAAI. However, apart from the studies presented above, providing stimuli before the scale has not been generally seen in recent studies concerning TAAI \cite{Zhang2023}.

\section{Materials \& Methods}
To address the issues mentioned above, aided by the guidelines of instrument development, we proposed the following steps to guide our development and validation of the teachers’ acceptance of AI instruments:

(1)	Define the constructs aiming to measure based on a theoretical framework.

(2)	Provide concise and clear stimuli at the beginning of the survey to assist participants in understanding AI and its applications in education.

(3)	Develop and revise items through expert reviews and think-aloud interviews. Each dimension includes multiple items to obtain variances. Item quality is also verified by item analysis.

(4)	Provide reliability and validity evidence using various methods. Expert reviews and teacher think-aloud procedures are used to ensure face and content validity. CFA and item analysis provide evidence of factor structure, reliability, convergent validity, discriminant validity, and item discrimination.

\subsection{Conceptualizing teachers’ acceptance of AI in education (TAAI)}
\subsubsection{Theories of technology acceptance }
Teachers’ acceptance of novel technologies has been one of the most critical topics in science and technology research. Teachers are the first exposers when technologies come into play in education. Regardless of the novelty of the technologies, teachers, knowledge facilitators, and class activity coordinators determine what technologies to adopt in classrooms, when, and how. Therefore, effective technology integration relies highly on teachers’ willingness and acceptance of the technology. Given the importance of adopting novel technologies, uncovering factors accounting for teachers’ acceptance of technology has become one of the crucial research topics within decades of literature. In this regard, Davis \cite{Davis1985} proposed the technology acceptance model (TAM) to explain factors influencing the acceptance of any novel technology, which has been the most influential model in the field. 

TAM concerns how users come to accept and use technology, including five factors that influence their decision about how and when they will use it. Davis underscores users’ perceptions, attitudes, behavioral intentions, and actual uses as integral constructs. He believed that teachers’ perceived usefulness (PU) and perceived ease of use (PEU) are at the forefront of other factors in their decisions to adopt the technology. Technologies with lower PU and PEU lose the first dip of being integrated into educational settings. Davis further pointed out that teachers’ attitude towards use (AU) plays a significant role in their decision-making, as negative feelings during technology integration can discourage further usage. Moreover, he realized that teachers’ reluctance to use technology is often because of the lack of intent to use it; the more teachers are self-motivated, the more likely they are to accept the technologies. Therefore, Davis perceived behavioral intention (BI) as one of the critical factors as well. Ultimately, Davis suggested that teachers have to use the technology in person to realize the potential of the technology; the more frequently teachers actually use (AU) the technology, the more likelihood that teachers may accept and integrate the technology in their classroom. 

In their later work, Davis and colleagues removed the construct of AU because of its generality and added two additional constructs: social influence processes (e.g., subjective norm) and cognitive instrumental processes (e.g., job relevance) \cite{venkatesh2000theoretical}. In addition, since a limited effect was found on the AU factor, many studies that applied the TAM have left out AU from the construct \cite{hu2003examining, park2009analysis, SanchezPrieto2017, tan2014predicting}.  Researchers proposed revised TAM models without the AU factor, such as TAM 2 \cite{venkatesh2000theoretical}, TAM 3 \cite{venkatesh2008technology}, and UTAUT \cite{venkatesh2003user}. Despite the broad use of TAM, the direct application of TAM in teachers’ acceptance of AI remains an area with limited research.

\subsection{Constructs of TAAI}
AI shares common features with prior learning technologies while maintaining its distinct characteristics, such as stimulating human-like cognitive functions \cite{Stahl2018}, presenting teachers with excitement and unique challenges as they embrace AI in teaching and learning settings. These challenges, in particular, are stressed by the automaticity, accessibility, and functionality of AI \cite{Alasadi2023}, which substantially impact teachers’ willingness to adopt AI in teaching and learning. 

AI seldom acts in isolation in the classroom; instead, it is integrated with other technologies to facilitate teaching and learning. For example, Gerard and Linn (2022) \cite{Gerard2022} included AI-based automatic scoring in web-based inquiry to facilitate science learning. Latif and Zhai, et al. \cite{Latif2024} included AI-Scorer in mobile learning to facilitate formative assessment practices in classrooms. These integrations increase the performance challenges for teachers due to the complexity of the applications, thus drawing teachers’ concerns about ease of use. What made it more complex was the unique applicability of machine algorithms for specific purposes and populations based on the training dataset. Research has suggested that machine algorithms trained by given datasets are expected to be used with new datasets with similar features to maintain objectivity and accuracy; violating this rule may result in errors or bias \cite{Zhai2023}. This requirement is substantially demanding as teachers are supposed to understand the conditions of AI and its applicability. Will teachers be able to use AI in a feasible manner? To what degree does this additional complexity impact teachers’ willingness to use AI? These are questions that TAAI has to address.

Although AI has recently received significant attention, many people are unaware of AI applications. According to Gartner Research Circle and Gartner’s 2019 CIO Agenda survey, 42\% of respondents did not fully understand the benefits of implementing AI in the workplace and daily life \cite{Gartner2019}.  In the educational field, although much research has been conducted on integrating AI into teaching practice, teachers’ actual use of AI in classrooms is still very limited \cite{AlKanaan2022}. Both in-service and pre-service teachers barely have experience learning AI knowledge and using AI technology \cite{AlKanaan2022}. Due to the circumstances, teachers’ perceived ease of use and perceived usefulness of AI can be important variables that impact their intention to accept AI. Thus, we adopt perceived ease of use and perceived usefulness as internal factors to explain teachers’ behavioral intentions. These three constructs are also empirically verified to be powerful for predicting and explaining user behaviors for new technologies in educational contexts \cite{Davis1989, SanchezPrieto2017, Zhang2023}. As a result, we have perceived ease of use for AI, perceived usefulness of AI, and behavioral intention of AI integration.

In addition, we expanded the TAM by adding two more constructs (i.e., anxiety and self-efficacy) to assess teachers’ acceptance of AI in education. Teachers’ anxiety about AI is critical to be assessed and garners significant scholarly attention \cite{hopcan2023exploring, Li2020, wang2022development}. The characteristics of AI, notably its ability to learn, reason, solve problems, and make decisions by imitating human cognitive abilities, compared to other information technologies \cite{Stahl2018}, raise critical concerns among teachers. One of the concerns is that the decision process of AI is not transparent and can potentially be biased, which makes it possibly to lead to unfair treatment of students \cite{nguyen2023ethical}. Some educators also worry that AI can cause job replacement, resulting in job loss and a decline in the quality of education \cite{terzi2020adaptation}. At the same time, since the functioning of AI applications is based on large and processing data, personal privacy also becomes an issue. With these concerns and the lack of trust in using AI in teaching, it will be hard for teachers to engage in the open and effective integration of AI in education \cite{hopcan2023exploring}. 

Besides, anxiety has been considered one of the important variables in predicting acceptance of AI. Many studies in educational contexts have explored the relationship of anxiety with other constructs, including attitudes, perceived ease of use, etc., and have found some significantly negative correlation \cite{Chen2012, hopcan2023exploring, SanchezPrieto2017, Wang2021}. At the same time, although teachers’ AI anxiety is increasingly investigated, existing research uncovered limited evidence of the relationship between teachers’ anxiety and their acceptance of AI. In this study, we proposed that teachers’ anxiety about using AI in teaching is a critical construct in the TAAI instrument. 

Teachers’ self-efficacy refers to teachers’ belief in their ability to integrate new technologies into their professional practice to enhance students’ learning \cite{SanchezPrieto2017}. Previous studies have found both direct and indirect effects of self-efficacy on pre- and in-service teachers’ acceptance of using new technologies, including mobile devices \cite{SanchezPrieto2017}, computers \cite{wong2012influence}, assistant technologies for special education \cite{nam2013acceptance}, etc. Self-efficacy, thus, is regarded as one of the key drivers in the adoption of AI \cite{Wang2024}. Due to the low actual use of AI in current teaching practice, teachers are somehow unfamiliar with learning and teaching with AI applications. They may also have a low level of awareness about the use of AI.  In fact, most teachers consider themselves unknown to the general characteristics of AI and how to apply it in teaching \cite{AlKanaan2022, incerti2020preservice}. The uncertainty can make teachers feel overwhelmed about using AI \cite{Wang2024} and undermine their perceived usefulness of adopting AI in teaching \cite{Wang2024}. Low self-efficacy in using AI in teaching, thus, leads to unwillingness to integrate AI into classroom practices. In this study, we also examined the teachers’ self-efficacy in using AI.
Given the advantages of TAM and its wide application in related studies, this study proposed the constructs of interest—TAAI within a revised framework of TAM. 

Table I presents the definitions of the construct through five sub-dimensions.
Table I. Constructs and descriptions of the instrument dimensions of teachers’ acceptance of AIED

\begin{table*}[]
\caption{Constructs and descriptions of the instrument dimensions of teachers’ acceptance of AIED}
\begin{tabular}{lp{15cm}}
\hline
\multicolumn{1}{c}{Dimension} & \multicolumn{1}{c}{Description}                                                                                                                            \\ \hline
Behavioral Intention          & The strength of teachers’ intention to use AI in their teaching practice                                                                                   \\
Behavioral Intention          & The degree to which teachers believe that AI can be used effortlessly                                                                                      \\
Perceived Usefulness          & The extent to which teachers feel that utilizing AI will improve their teaching performance                                                                \\
Self-efficacy                 & The degree to which teachers believe that they have the ability to perform specific tasks using AI in their teaching to achieve better educational results \\
Anxiety                       & Negative feelings and concerns that teachers can have when they use AI in their teaching practices                                                         \\ \hline
\end{tabular}
\end{table*}

\subsection{Development of the initial instrument }
The initial version of the TAAI instrument includes three sections. The first section consists of reading material and 10 five-point scale questions. The reading material introduces the concept of AI, some daily applications, and how AI is integrated into educational settings. The questions ask participants to illustrate their previous experiences of using AI applications in their personal and professional lives. The reading material and the questions function as the stimulus in the instrument to help participants familiarize themselves with AI applications in their daily lives and education settings. The second section collects participants’ demographic information, including gender, grade, and majors, to better contextualize students’ responses to the following questions. The third section includes 32 survey items to assess the five dimensions of teachers’ acceptance of AI. The survey items were developed using a five-point Likert scale, ranging from strongly disagree (1) to strongly agree (5).

The construction of teachers’ AI acceptance guided the development of the item pool in the third section of the instrument. We developed the items based on previous technology acceptance instruments, mostly for teachers \cite{Chocarro2023, Li2020, SanchezPrieto2016, SanchezPrieto2017, wang2022development, Wang2021}. Since some items focused on general AI acceptance rather than TAAI, we revised them to make the context more related to teachers’ using AI in their teaching. For example, the original item, “AI treats different people differently, which makes me anxious” \cite{Li2020}, was revised to form the item, “I am concerned that AI algorithms may be biased, leading to varying accessibility to students with diverse backgrounds” in AN dimension. Meanwhile, we adopted some items from other non-AI instruments to be suitable in the TAAI context. For instance, the item “Using WBLS saves me time” \cite{wang2009empirical} was modified into “Using AI technologies saves me time.” Additionally, we created new items focusing on teachers’ potential teaching practice with AI, such as the items “I am willing to use AI tools to help prepare lessons” and “I am willing to use AI tools to assess students,” which were developed as original items in the BI dimension to focus more on using AI for different teaching purposes. 

To establish the face and content validity  of the initial instrument, a panel of four experts —comprising two professors with specialization in AI in education, one with a focus on STEM education, and another with expertise in none-STEM education — conducted a comprehensive evaluation. They evaluated every part of the instrument concerning its intended purpose, phrasing, and the defined scope of each item’s corresponding dimensions. Furthermore, they provided insightful comments and suggestions for instrument revision at item and dimension levels. Based on expert feedback, we revised the instrument. We reduced the length of the reading material to make it more concise, clear, and readable. As for teachers’ experience of AI, we provided more examples of commonly used AI programs to make the items more familiar to respondents. For example, in the original item “I employ personal assistants in the mobile phone”, we added Siri as an AI assistant example that people usually use. For items that measure TAAI, we revised the wording of some items and also added concrete examples to make the items easier to read and understand. For example, we added “lesson preparation, grading, and other administrative tasks” as ways that teachers can use AI in teaching to save time. In total, 15 items in the instrument were modified.

\subsection{Think aloud}
To further establish face and content validity, we conducted think-aloud of the initial instrument with five pre-service teachers, including four females and one male, randomly selected from a class. Think-aloud protocol requires participants to articulate their feelings, thoughts, actions, and other cognitive processes while engaging in a set of activities to make thought processes as explicit as possible throughout task performance \cite{wolcott2021using}. This allows researchers to gain insights into the participants’ cognitive processes rather than just their responding results for further instrument refinement \cite{pepper2018think}. 
Researchers provided guidance for relevant concepts, functions, and procedures of think-aloud before the test, followed by a demonstration. The participants read the guidance and then completed a short task with think-aloud, ensuring their understanding of the procedure. Afterward, they were given the questionnaire and started to think aloud. Researchers videotaped the entire think-aloud process. Based on the data obtained, we modified or deleted the items that did not function as expected. For example, for some original items wording as “I can use AI technologies to …,” participating teachers commented that “Technology seems difficult. I can use AI-based tools, but I don’t think I can use AI technology since I know nothing about that.” We thus rephrased “AI technologies” as “AI applications” or “AI tools.” Through the think-aloud procedure, six items were revised, and an additional item was added. The added item is “I have seen others using AI tools for teaching” in the AI using experience section, since a participating teacher mentioned that although she had no experience of using AI in teaching, she has seen other teachers used AI applications.

\subsection{Field test and data collection}
Using a random sampling approach, we distributed the revised version of the TAAI instrument to 301 pre-service teachers for field test through an online platform called “Wenjuanxing.” All participants were from a four-year university specializing in in-service and pre-service teacher education. To increase the diversity of participants, we recruited participants with various majors, including STEM majors such as mathematics, physics, biology, chemistry, geography, and technology, and non-STEM majors including Chinese, English, history, and politics. Among the 301 participants, we collected 274 valid responses for data analysis. The demographic distribution of the sample is in Table II.

\begin{table}[]
\caption{Distribution of the sample according to gender, grade, and subject domain}
\resizebox{\linewidth}{!}{
\fontsize{11}{12}\selectfont
\begin{tabular}{lcccccc}
\hline
                      & \multicolumn{2}{c}{Gender} & \multicolumn{2}{c}{Grade}               & \multicolumn{2}{l}{Subject domain}        \\
                      & Male        & Female       & Undergraduate & Graduate                & STEM & \multicolumn{1}{l}{Non-STEM major} \\ \cline{2-7} 
\multicolumn{1}{c}{n} & 54          & 220          & 114           & \multicolumn{1}{l}{160} & 180  & 94                                 \\
\%                    & 19.7        & 80.3         & 41.6          & 58.4                    & 65.7 & 34.3                               \\ \hline
\end{tabular}
}
\end{table}
\subsection{Data analysis}
To examine individual items’ discrimination power, we first conducted a t-test using SPSS to compare teachers’ scores between the upper and lower 27\%. Meanwhile, using SPSS, we also calculated Cronbach’s alpha for the whole instrument, each dimension, and each item, to indicate the internal consistency of the scale and identify items that need to be deleted.

Then, we employed Mplus to conduct confirmatory factor analysis (CFA) to confirm the dimensionality of the instrument, providing evidence for construct, convergent, and discriminate validity. We used the weighted least squares means and variance adjusted (WLSMV) estimator option, considering the categorical and non-normal data.  Factor structure was verified based on a solid theoretical foundation and by conducting CFA using empirical data. CFA can help examine whether the data fit a theoretical structure. Based on the results of CFA, many items were loaded onto the factors; however, the fit of the overall model was not satisfying enough. Thus, according to CFA results, including factor loadings and modification indices (M.I.), we removed ill-fitting items with factor loadings less than 0.32 (Tabachnick \& Fidell, 2001). Meanwhile, factor loadings of the items belonging to the same dimension should be close ideally, and items with significantly lower factor loadings were also considered to have relatively poor performance. Furthermore, items with M.I. larger than 10 were “bad” \cite{muthenbo} since M.I. shows the degree to which model fit improves if an item is allowed to be correlated with another factor or removed. 

To select and exclude ill-items, we considered item complexity and content. For example, the item “Learning to use AI in teaching is difficult for me” showed relatively low factor loading (i.e., 0.44) and a high modification index (i.e., above 10). Although it stands for the “learning anxiety” construct in anxiety construct, the concept can be easily confused with “self-efficacy,” which can also be seen from the think-aloud result. This will result in a correlation between multiple factors for one item. Therefore, we decided to remove the item. Another example is the item “I will actively learn to adopt AI tools to assist teaching.” The factor loading was satisfying (i.e., 0.896) while the M.I. was above 10, the meaning expressed in the item is somewhat different from other items in the dimension, which concerned teachers’ willingness to use AI in teaching practice. Therefore, we also removed the item. On the contrary, for the item “I would like to use AI tools for student assessment” in the behavioral intention dimension, the factor loading (i.e., 0.741) was acceptable, while the M.I. was above 10. We decided to keep the item since assessment is an important part of teaching practice. The cull of items stopped until the overall fit of the model and the statistics of each item reached a satisfactory result.

\section{Results}
The final instrument contains five sub-dimensions and 27 items in total. We reported the psychometric quality results of the final instrument as follows.

\subsection{Item analysis}
To examine the item discrimination, we conducted a two-sample t-test to compare scores between the upper 27\% and the lower 27\% of participants for each item \cite{Yilmaz2023}. We found that all items show significant discrimination between the upper and lower groups of participating teachers (see Table III). The findings suggest the robustness of individual items in discriminating participants with varying levels of acceptance of AI.


\begin{table*}[]
\caption{Item analysis results}
 \centering
\resizebox{\linewidth}{!}{
\begin{tabular}{ccccccc}
\hline
\multirow{2}{*}{Item} & \multirow{2}{*}{Cronbach’s alpha if item deleted} & \multicolumn{2}{c}{Mean (SD)} & \multirow{2}{*}{t} & \multirow{2}{*}{Upper 27\%} & \multirow{2}{*}{Lower 27\%} \\ \cline{3-4}
                      &                                                   & Upper 27\%     & Lower 27\%     &                       &                              &                              \\ \hline
\multicolumn{7}{l}{PU} \\ 
Y1                    & 0.91                                              & 4.33 (0.54)    & 3.59 (0.58)    & -8.996*              & 4.33                         & 3.59                         \\ 
Y2                    & 0.92                                              & 4.29 (0.58)    & 3.43 (0.77)    & -8.488*              & 4.29                         & 3.43                         \\ 
Y3                    & 0.92                                              & 4.33 (0.58)    & 3.54 (0.69)    & -8.379*              & 4.33                         & 3.54                         \\ 
Y4                    & 0.91                                              & 4.28 (0.63)    & 3.38 (0.77)    & -8.685*              & 4.28                         & 3.38                         \\ 
Y5                    & 0.92                                              & 4.36 (0.59)    & 3.72 (0.75)    & -6.487*              & 4.36                         & 3.72                         \\ 
Y6                    & 0.92                                              & 4.40 (0.49)    & 3.75 (0.60)    & -8.019*              & 4.40                         & 3.75                         \\ 
\multicolumn{7}{l}{PEU} \\ 
Y7                    & 0.92                                              & 3.88 (0.71)    & 2.78 (0.75)    & -10.253*             & 3.88                         & 2.78                         \\ 
Y8                    & 0.92                                              & 3.70 (0.77)    & 2.49 (0.67)    & -11.462*             & 3.70                         & 2.49                         \\ 
Y9                    & 0.91                                              & 3.64 (0.70)    & 2.50 (0.69)    & -11.228*             & 3.64                         & 2.50                         \\ 
Y10                   & 0.92                                              & 3.85 (0.65)    & 2.73 (0.87)    & -10.194*             & 3.85                         & 2.73                         \\ 
Y11                   & 0.91                                              & 3.90 (0.63)    & 2.48 (0.70)    & -14.469*             & 3.90                         & 2.48                         \\ 
\multicolumn{7}{l}{BI} \\ 
Y12                   & 0.91                                              & 4.36 (0.53)    & 3.53 (0.78)    & -8.447*              & 4.36                         & 3.53                         \\ 
Y13                   & 0.91                                              & 4.35 (0.52)    & 3.18 (0.89)    & -10.820*             & 4.35                         & 3.18                         \\ 
Y14                   & 0.92                                              & 4.14 (0.72)    & 3.12 (1.04)    & -7.768*              & 4.14                         & 3.12                         \\ 
Y15                   & 0.91                                              & 4.38 (0.49)    & 3.27 (0.92)    & -10.253*             & 4.38                         & 3.27                         \\ 
Y16                   & 0.91                                              & 4.37 (0.53)    & 3.51 (0.79)    & -8.654*              & 4.37                         & 3.51                         \\ 
\multicolumn{7}{l}{SE} \\ 
Y17                   & 0.91                                              & 4.21 (0.50)    & 3.39 (0.76)    & -8.487*              & 4.21                         & 3.39                         \\ 
Y18                   & 0.91                                              & 4.18 (0.53)    & 3.11 (0.80)    & -10.508*             & 4.18                         & 3.11                         \\ 
Y19                   & 0.91                                              & 4.23 (0.49)    & 3.27 (0.71)    & -10.581*             & 4.23                         & 3.27                         \\ 
Y20                   & 0.91                                              & 4.15 (0.53)    & 3.10 (0.76)    & -10.924*             & 4.15                         & 3.10                         \\ 
Y21                   & 0.91                                              & 4.05 (0.62)    & 2.92 (0.82)    & -10.656*             & 4.05                         & 2.92                         \\ 
Y22                   & 0.91                                              & 4.32 (0.54)    & 3.21 (0.67)    & -11.228*             & 4.32                         & 3.21                         \\ 
\multicolumn{7}{l}{AN} \\ 
Y23                   & 0.92                                              & 2.72 (1.05)    & 2.03 (0.70)    & -4.964*              & 2.72                         & 2.03                         \\ 
Y24                   & 0.92                                              & 2.80 (1.02)    & 2.26 (0.89)    & -3.857*              & 2.80                         & 2.26                         \\ 
Y25                   & 0.92                                              & 2.78 (1.16)    & 2.10 (0.85)    & -4.572*              & 2.78                         & 2.10                         \\ 
Y26                   & 0.92                                              & 2.50 (1.00)    & 2.12 (0.71)    & -2.977*              & 2.50                         & 2.12                         \\ 
Y27                   & 0.92                                              & 2.52 (1.02)    & 1.93 (0.63)    & -4.700*              & 2.52                         & 1.93                         \\ \hline
\multicolumn{7}{l}{*p $ < $ 0.01} \\ 
\end{tabular}
}
\end{table*}

\subsection{Reliability}
To examine the reliability of the instrument, we calculated Cronbach’s alpha for the instrument, each dimension, and each item if deleted (see Table III). Results indicate that the reliability coefficient of the instrument is 0.92, and that of the sub-dimensions PU, PEU, BI, SE, and AN are 0.88, 0.91, 0.91, 0.91, and 0.77, respectively. All the values over 0.7 indicate a high consistency in each dimension and among the instrument \cite{N}(Nunnally, 1978). Meanwhile, we found that if an item was deleted, the Cronbach’s alpha was either dropped or similar to the original value, which means no item should be deleted.

\subsection{Validity}

\subsubsection{Construct validity}
To ensure the validity, CFA was utilized to evaluate how well the measurement model fit the data by reporting root mean square error of approximation (RMSEA), comparative fit index (CFI), Tucker-Lewis index (TLI), a Standardized Root Mean Square Residual (SRMSR), and chi-square ratio on the degrees of freedom ($\chi^2/\text{df}$). RMSEA is an absolute fit measure that is one of the most widely used indices in SEM \cite{kline2023principles}(Kline, 2015) with a value less than 0.08 is acceptable \cite{Byrne2010}. CFI and TLI are incremental fit measures, ranging from 0 to l, with a value greater than 0.90 suggests a good fit \cite{Hair1998}. SRMSR is an absolute fit measure, ranging from zero to 1.0 with a value less than 0.07 being acceptable \cite{hu1999cutoff}. As for χ2/df, it is preferred to chi-square fit statistic since it is not affected by large samples. A value less than 3 indicates a good fit between the hypothesized model and the sample data (Chi-square for model fit in confirmatory factor analysis). The results of CFA showed a good fit of RMSEA of 0.061, a CFI of 0.981, a TLI of 0.979, an SRMSR of 0.051, a chi-square of 629.186 with df =314, and a χ2/df of 2.00. The test results indicated that the instrument’s items clustered around factors or sub-dimensions that represented the main components as constitute the theoretical framework. The standardized item loading ranges within the five factors are perceived usefulness (0.70-0.94), ease of use (0.83–0.94), behavioral intention (0.75–0.95), self-efficacy (0.81–0.92), and anxiety (0.60–0.81) (see Table IV), which are all above the benchmark of 0.5 \cite{Wang2024}.

\begin{table}[]
\caption{Standardized loadings for teachers’ acceptance of AI}
\centering
\begin{tabular}{llllll}

\hline
Item   & PU     & EU     & BI     & SE     & AN     \\ \hline
Y1   & 0.94 &      &      &      &      \\
Y2   & 0.81 &      &      &      &      \\
Y3   & 0.80 &      &      &      &      \\
Y4   & 0.88 &      &      &      &      \\
Y5   & 0.70 &      &      &      &      \\
Y6   & 0.89 &      &      &      &      \\
Y7   &      & 0.83 &      &      &      \\
Y8   &      & 0.88 &      &      &      \\
Y9   &      & 0.89 &      &      &      \\
Y10  &      & 0.86 &      &      &      \\
Y11  &      & 0.94 &      &      &      \\
Y12  &      &      & 0.90 &      &      \\
Y13  &      &      & 0.95 &      &      \\
Y14  &      &      & 0.75 &      &      \\
Y15  &      &      & 0.91 &      &      \\
Y16  &      &      & 0.93 &      &      \\
Y17  &      &      &      & 0.81 &      \\
Y18  &      &      &      & 0.86 &      \\
Y19  &      &      &      & 0.91 &      \\
Y20  &      &      &      & 0.89 &      \\
Y21  &      &      &      & 0.81 &      \\
Y22  &      &      &      & 0.92 &      \\
Y23  &      &      &      &      & 0.75 \\
Y24  &      &      &      &      & 0.69 \\
Y25  &      &      &      &      & 0.69 \\
Y26  &      &      &      &      & 0.60 \\
Y27  &      &      &      &      & 0.81 \\ \hline
\end{tabular}
\end{table}

\subsubsection{Convergent validity }
The convergent validity was tested using factor loadings, Average Variance Extracted (AVE), and composite reliability (CR). Based on previous research, the factor loading should exceed 0.60 to demonstrate the sufficiency of representing the construct \cite{molefi2023using}. Findings suggest that our item loadings meet the benchmark (see Table IV). Furthermore, Table V shows the AVE and CR of each factor. AVE measures the amount of variance captured by a latent variable from its measurement scale to the amount of variance due to measurement errors \cite{Fornell1981}, while CR assesses how well a construct is measured by the indicators assigned to it. The AVE of all factors is above 0.5, indicating that the latent variables account for at least 50\% of the measurement variance, which stands for a satisfying convergent validity \cite{Fornell1981}. The value of CR on each factor is above 0.8, indicating a robust composite reliability \cite{Hair2011}.

\begin{table}[]
\caption{AVE and CR of each dimension }
\centering

\begin{tabular}{llllll}
\hline
Indicators & \multicolumn{5}{c}{Factors}      \\ \hline
           & PU   & EU   & BI   & SE   & AN   \\
AVE        & 0.70 & 0.78 & 0.66 & 0.76 & 0.51 \\
CR         & 0.93 & 0.95 & 0.91 & 0.95 & 0.83 \\ \hline
\end{tabular}
\end{table}

\subsubsection{Discriminate validity}

To examine the discriminate validity, we calculated the correlations between the dimensions (see Table VI). According to the correlation results, most of the dimensions were statistically correlated to each other, with a value higher than 0.4, indicating a common construct across them. However, the dimension anxiety was not significantly correlated with perceived usefulness and perceived ease of use ($p\leq 0.01$) but was correlated with behavioural intention as well as self-efficacy significantly with a relatively low correlation coefficient.

Additionally, we compared the correlation coefficients between factors and the Average Variance (AV) of each construct. AV is the square root of the corresponding AVE. When the AV of each factor is more than the correlation coefficients of that dimension with other constructs, it indicates a valid discriminant \cite{chin1998partial}. Table VI shows the correlation between constructs, which was smaller than AV, indicating robust discriminant validity.

\begin{table}[]
\caption{Correlation between factors}
\resizebox{\linewidth}{!}{
\begin{tabular}{llllll}
\hline
\multicolumn{1}{c}{Factors} & PU     & PEU    & BI      & SE      & AN    \\ \hline
PU                          & 0.838  & —      & —       & —       & —     \\
PEU                         & 0.430* & 0.884  & —       & —       & —     \\
BI                          & 0.835* & 0.430* & 0.815   & —       & —     \\
SE                          & 0.652* & 0.722* & 0.621*  & 0.870   & —     \\
AN                          & -0.106 & -0.082 & -0.146* & -0.222* & 0.711 \\ \hline

\multicolumn{6}{c}{*$p< 0.01$; The AV of each dimension on the diagonal line.}
\end{tabular}
}
\end{table}

\section{Discussion and Conclusion}
Promoting teachers’ integration of AI applications in classrooms is vital in the AI era, while this classroom innovation depends on teachers’ acceptance of AI in education. Since teachers’ acceptance will impact their willingness and ways to adopt AI in teaching, it is critical to understand teachers’ acceptance of AI. Due to the lack of well-validated instruments measuring teachers’ acceptance of AI, this study developed and validated a high-quality instrument to bridge the gap in the literature. The development of the instrument in this study addresses concerns that have been found in previous instruments. We used multiple methods to provide various evidence for the instrument’s psychometric quality, including using expert review and think-aloud interviews to provide evidence for face and content validity, using robust statistical measures (i.e., CFA) to explore the underlying psychometric properties and the structure of the instrument, including construct, convergent, discriminant validity and reliability. The final instrument had five dimensions with 27 items, consistent with the proposed theoretical framework, with a satisfying model fit, reliability, and validity. According to Potvin \& Hasni (2014) \cite{potvin2014interest}, such empirical alignment between the factor structure and item loading of the instrument and the theoretical basis is also important evidence of the instrument’s construct validity. The final instrument items is in the Appendix, including sub-dimensions as Behavioral Intention of AI (five items), Perceived Ease of Use AI (five items), Perceived Usefulness of AI (six items), Self-efficacy of AI (six items), and AI Anxiety (five items). The various pieces of evidence suggest that the instrument is valid and reliable for measuring teachers’ acceptance of AI in education.

The TAAI instrument has significant implications for promoting AI uses in classrooms and teacher professional development. For example, future research can use the instrument to evaluate pre-service teachers’ acceptance of using AI in teaching and provide evidence for relative intervention in teacher education and professional development. Furthermore, it can be used to further explore additional influential factors correlated to the constructs in this instrument to deepen the understanding of teachers’ AI acceptance in education, such as gender, AI-related experience, and cultural backgrounds. Some of these variables, such as gender, have been paid attention to in teachers’ attitudes and application of AI \cite{Alissa2023,hopcan2023exploring,Zhang2023}, while others have been explored in users’ acceptance of other new technologies in other contexts \cite{kelly2023factors}. However, little research has been conducted to investigate the influence of those external factors on teachers’ acceptance of AI. Moreover, future research can explore the causal relationship between behavioral intention and other constructs of the TAAI instrument to explain how these factors could impact teachers’ decision-making and behavior in using AI in teaching. 

We acknowledge some limitations for future studies to address. First, although the participating pre-service teachers came from diverse disciplines, they were from one university, which could limit the generalization of the instrument. Future studies should recruit more participants in other settings to further validate the instrument. This will contribute to the refinement of the instrument according to data obtained from more teacher groups. Another limitation is the country-specific context. While AI in education is popular around the world, the specific applications in the classroom may be different. To cater to the Chinese context, the examples we provided about AI tools in the stimulus of the TAAI were customized in China. To use the instrument in other cultural contexts, researchers should make some revisions of the AI application examples according to their specific contexts. The revised instrument can allow comparison studies to explore the possible characteristics and cross-cultural differences in teachers’ acceptance of AI worldwide. 

\section*{Conflict of Interest}
No potential conflict of interest was reported by the authors.

\section*{Acknowledgment}
The authors thank China Scholarship Council (CSC) for supporting the study. Any opinions, findings, conclusions, or recommendations expressed in this material are those of the author(s) and do not necessarily reflect the views of the CSC.


\section*{Ethics approval statement}
All procedures performed in studies involving human participants were in accordance with the ethical standards of the institutional and/or national research committee and with the 1964 Helsinki declaration and its later amendments or comparable ethical standards. This article does not contain any studies with animals performed by any of the authors.

\bibliographystyle{ieeetr}
\bibliography{ref}

\end{document}